# Creation and Evaluation of Software Teams - A Social Approach

**Surayne Torres*, Yadenis Piñero and Pedro Piñero**
Departamento de Gestión de Proyectos.
Universidad de las Ciencias Informáticas (UCI), Carretera de San Antonio de los Baños.
Km 2 ½. Torrens. La Habana. Cuba.
Teléfono: + 537 837 2409.
E-mail: storres@uci.cu
E-mail: ppp@uci.cu
E-mail: yadenispinero@gmail.com
*Corresponding author

**Luiz Fernando Capretz**
Department of Electrical and Computer Engineering
Faculty of Engineering
The University of Western Ontario
London, Ontario.
E-mail: lcapretz@uwo.ca

## Abstract

This work discusses an important issue in the area of human resource management by proposing a novel model for creation and evaluation of software teams. The model consists of several assessments, including a technical test, a quality of life test and a psychological-sociological test. Since the technical test requires particular organizational specifications and cannot be examined without reference to a specific company, only the sociological test and the quality of life tests are extensively discussed in this work. Two strategies are discussed for assigning roles in a project. Initially, six software projects were selected, and after extensive analysis of the projects, two projects were chosen and correctives actions were applied. An empirical evaluation was also conducted to assess the model's effectiveness. The experimental results demonstrate that the application of the model improved the productivity of project teams.

## Introduction

There are various techniques and guidelines for improving the process of building project teams. However, these guidelines should be adapted to specific environments. Generally, each member of a given team possesses a special area of expertise or natural ability that should be utilized by project managers. Accordingly, many successful organizations depend on the optimal mix of competence, trust and mutual esteem in team relationships.

Human resource management is an interdisciplinary area in project management. Some project managers perceive and manage individuals as if they were modular components rather than unique team members; however, software production processes are different from other industrial production processes. During software production, many problems that occur are directly related to software teams and to the mutual relationships among their members. For instance, (DeMarco & Lister, 1999) argue that team relationships are highly relevant, and consequently, there are four elements that affect human resources: the management of human resource techniques, human resource acquisition processes, activities that improve team productivity and the office environment. According to (Curtis et al, 1988), human resource selection and management is more important than technologies and tools. The IEEE vice-president suggests that in order to develop a successful project, managers should focus on understanding the project goals, appropriately handling the flow of ideas, and honing the team members' relationships (Weinberg, 1986). Overall, he maintains that the quality of products depends on software teams, where each member contributes to the quality by performing his/her part.

In general, selection processes consist of applying technical tests and interviews. However, these procedures alone do not ensure the selection of successful software teams, especially since interviews do not always properly account for all aspects of human behavior. As (DeMarco & Lister, 1999) explain, the skill tests are usually focused on the tasks that candidates would perform at the beginning of the work. However, these tests do not necessarily guarantee the correct evaluation of each candidate during the entire project. Members of a software team often change their activities or roles during the span of the project, thus indicating that such tasks have not been adequately considered during the initial human resource acquisition process.

Other viewpoints about the selection of software teams are presented by (Edgemon, 1995) and (Pressman, 2005); Edgemon proposes the following four areas: problem resolution, leader skills, reward management, and sociological behavior. There are several tests to assess the personality of individuals (Myers et al, 1985)(Catell et al., 2008)(Belbin & Mead, 2010). However, none has the particularity of evaluating people in normal situations and stress situations; this is an important element in the work environments of software development.

On the other hand, Pressman promotes project management on the basis of four elements, known as the four “Ps:” Personal, Product, Process and Project. The order of Pressman’s elements is not arbitrary, as he explicitly states that personal management is the most important aspect in software projects.

The Project Management Institute deals with human resource management, process organization, and the management and leadership of project teams. Accordingly, the Institute has proposed the following four processes: developing human resources plans, acquiring a project team, developing a project team, and managing a project team. There are four techniques for acquiring project teams, as described in the guide to the Project Management Body of Knowledge (PMBOK) (Project-Management-Institute, 2004): pre-assignment, negotiation, acquisition and virtual teams. Although the PMBOK guide is one of the most accepted international standards of project management, it constitutes an abstract guide that should be adapted to specific situations and particular environments.

The People Capability Maturity Model (P-CMM) defines “staffing” as one of the prime process areas at the “Managed Level” (Curtis, et al., 2009), thus indicating the importance of staffing for organizations. Specifically, the purpose of staffing is to establish a process where qualified individuals are recruited, selected, and transitioned into assignments. The “ability to perform” statements include the required definitions used in an organization’s selection process and the necessary methods and procedures for individuals involved in staffing activities. Moreover, the “practices” description establishes that a selection process and appropriate selection criteria are defined for each available position. In particular, Thomsett considers the team’s relationships highly relevant for a project’s success (Thomsett, 1990).

TSPi is a methodology that provides a defined process to develop software by teams. TSPi aim is to show defined process components (roles, scripts, forms and standards) (Carleton, A., et al., 2010). However, this methodology does not show how to form a good team.

This paper proposes a model that concerns the acquisition of human resources for software teams. The main idea of this model entails the combination of technical expertise and the sociological relationships among team members. This proposal can be utilized independently of the software development methodology used or the size of the team. In section 2 describes the techniques involved in the new model and provides details for acquiring the model's algorithm. However, these techniques are not an appropriate substitute for human expertise; rather, they solely constitute a decision-making tool. Section 3 analyses the experimental results, and Section 4 presents concluding remarks.

## A social model for acquiring software development teams

Technical knowledge is considered a prime requirement among software team members; however, elements pertaining to human resources also need to be considered. Specifically, these elements include sociological behavior and human relationships, technical knowledge and software team competencies, and the quality of life for software team members, as shown in Figure 1(a).

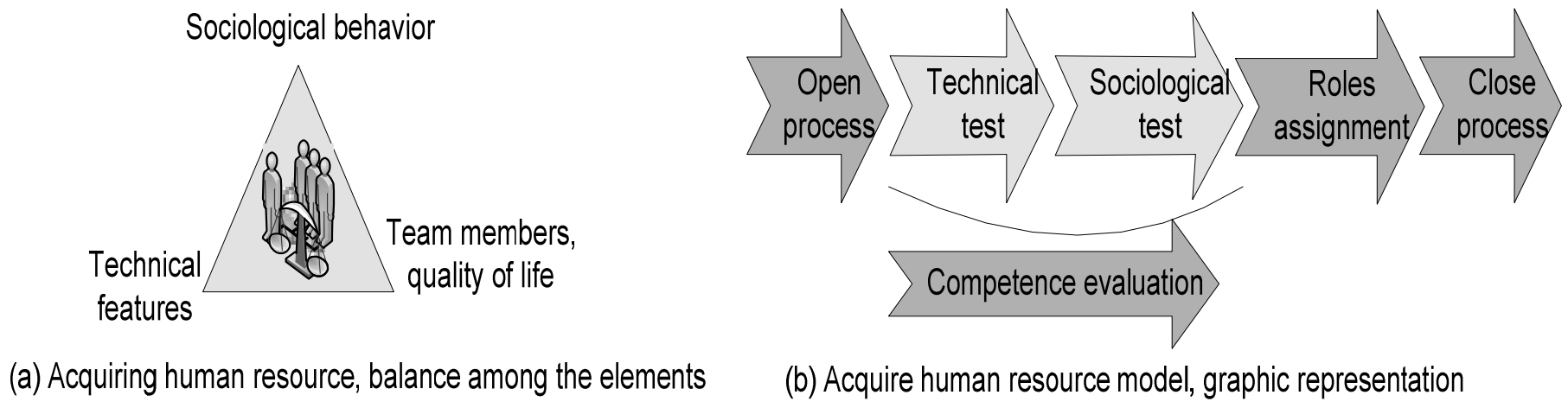

Figure 1 Graphical representation and balance among the elements.

Thus in order to achieve the optimal balance during the software development process, the human resource selection process should guarantee equilibrium among these elements.

The model presented in this paper consists of four processes, as depicted in Figure 1(b):

Process 1: Open process and initialization,

Process 2: Competence evaluation process and interviews,

Process 3: Roles assignment process; and,

Process 4: Close process.

### Open Process

Prior to the open process, team managers need to know the project objectives. Subsequently, the managers should define four milestones:

a) Create the human resources management group
b) Establish the number of work places in a human organization chart
c) Define the specific roles required for the project
d) Receive personal requests

In order to obtain the first milestone, the project manager should create a special group, the HR management group, or they should contact human resources management services for outsourcing.

For the second and third milestones, human resource management experts should define a hierarchical organization chart.

The fourth milestone consists of a voluntary request list, which requires the candidate's name, contact address, possible role, and other basic information. By the end of this step, project managers should have a list of candidates interested in the project.

**Competence Evaluation Process**

The proposed model recommends the application of three aptitude tests to each candidate: a technical test, a sociological test and a quality of life test. First, the technical test should be applied in conjunction with each candidate's role aspirations and should be based on the competency evaluation processes. As previously mentioned, each organization should define the required roles by considering the characteristics of the team members. The technical test should be developed according to these requirements. For example, in software production projects, the common roles include analyst, designer, architect, developer and project manager. However, in the technical test, the roles are entirely dependent on the project features. Accordingly, (Brainbench Previsor Company, 2008) and (Verio, 2008) have discussed test solutions for technical skills.

The second test consists of a questionnaire for evaluating the sociological state of candidates (Aragon, 2007). This assessment provides an integrated perspective of individuals' conduct under normal conditions as well as in tense situations. Specifically, the test evaluates candidates' activity level in a group and their attitudes towards people in a work environment. As a result, project managers can utilize these tests to predict an individual's personal behavior prior to their assignment in a software project. There are two elements in this proposed test: a sociological questionnaire (Tables I and II) and a guide for applying it (Section 1).

**1) The Sociological Questionnaire**

The questionnaires presented in Tables 1 and 2 have been created for this work based on (Gomez & Acosta, 2003). The third group of tests consists of a questionnaire for evaluating the quality of life of candidates, as presented in Table 5.

Table 1 Sociological questionnaire 1 for human resources evaluation

| **QUESTIONNAIRE 1** | |
|---|---|
| I like to act… | If I am in disagreement |
| A. Friendly and support other people. | a. I appeal to the sense of justice and legality of other people. |
| B. Quickly and decisively with others. | b. I try to be smarter and maneuverable. |
| C. Compact and firm with others. | c. I stay quiet. |
| D. As appropriate every time. | d. Try again and/or open a new point of view. |
| I frequently try to be….. | When I fail…. |
| E. Modest and idealist. | e. I feel panic and look for others to support me. |
| F. Persuasive and winner. | f. I keep on pushing because of my ideas. |
| G. Patient and realistic. | g. I remain quiet and inflexible. |
| H. Nice and real. | h. I keep my mind open and I continue joyfully. |
| People see me as … | People who look at me in my worst moments, say I am … |
| I. A trustful and advisable person | i. Humble and emotional. |
| J. A self-confident person who takes the initiative and acts. | j. Aggressive and commanding. |
| K. A careful, conscious and a systematic person. | k. Stubborn/bull-headed and absent minded. |
| L. An enthusiastic person who understands easily and adapts to any situation. | l. Superficial/shallow and disloyal. |

Table 2 Sociological questionnaire 2 for the human resource evaluation

| QUESTIONNAIRE 2 | |
|---|---|
| Usually I want to…. | In times of stress, I ... |
| A. Move forward with pride to great ideals. | a. Assume more responsibilities and remain robust. |
| B. Take control of the situation and reach the goals. | b. I get impatient and act quickly. |
| C. Be systematic, logical and a sound thinker. | c. I prove what I say with real data and information. |
| D. Win the people being insistent and convincing. | d. I try not to interfere with others |
| I usually treat others.... | In moments of stress I relate to others... |
| E. By being polite. | e. Being gullible and easily influenced. |
| F. In an active way and focusing on tasks. | f. Being dominant and impulsive. |
| G. In a methodical manner. | g. Being shy and distrustful. |
| H. In a friendly way. | h. Being very flexible. |
| I want to see myself as ... | People see me sometimes as ... |
| I. A loyal and trustworthy person. | i. Having little confidence in myself. |
| J. A competent and active person. | j. Being a tough negotiator. |
| K. A careful and logical person. | k. Being stubborn and determined. |
| L. A flexible and comprehensive person. | l. Being inconsistent to attract attention. |

**The sociological test guide**

This section presents the steps for analyzing individual personalities. Additionally, it discusses some tools that analyze team balance in the sociological test, ensuring that software teams consist of diverse personalities that create equilibrium amongst team members and minimize discord.

The validation of the scales of the measures used in this test was performed through the application of the Delphi Method to 29 experts from different organizations of Cuban software, dedicated to the management or human resources research. There were three rounds where each one was evaluated by experts at different scales for the measurements. Four experts were eliminated so in the final round there were only 25. The stadigraphs that were used in the study was the mean, mode and standard deviation that give us an overview of the results obtained in each of the questions (Torres, 2011).

Step 1: Create a graph to represent an integrated view of a person's characteristics, as shown in Figure 2.

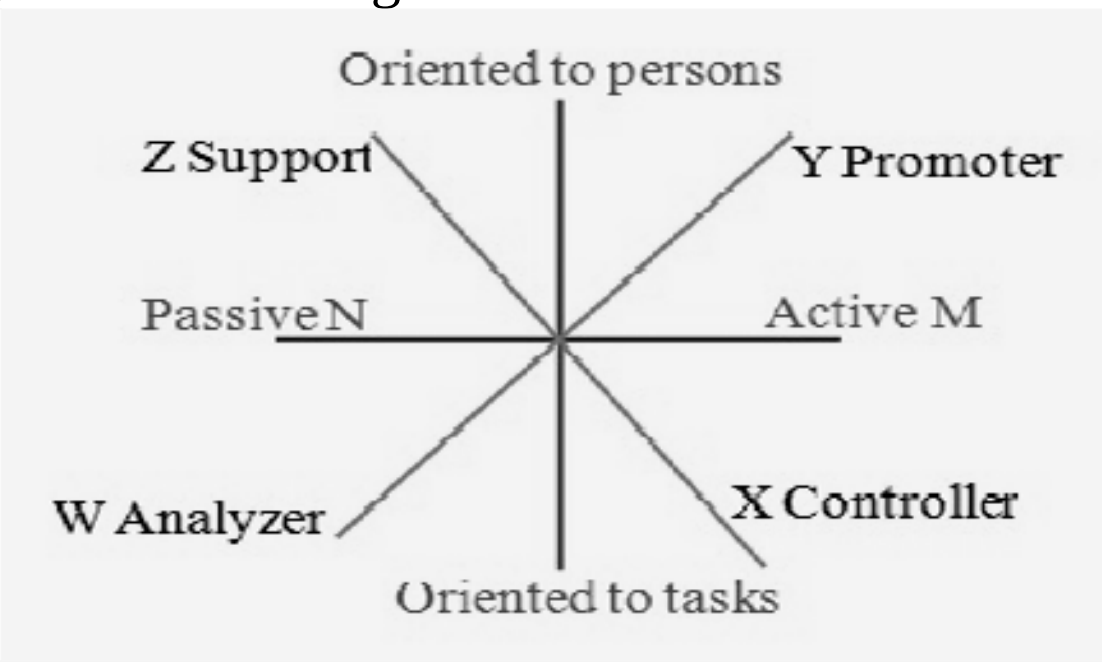

Figure 2 Chart to represent the features of each person

Step 2: Complete the questionnaires presented in Table 1 and Table 2. Each question has four possible answers to which respondents should assign a value between 1 and 4; repeated values are not permitted. Higher values mean that respondents believe they possess a certain characteristic, whereas lower values indicate that respondents do not associate themselves with a particular attitude.

Step 3: Summarize the results by using Equation 1 and 2.

Equation 1 and 2: Set of equations to summarize the questionnaire results.

$$\left.\begin{matrix} A+A+E+E+I+I=Z \\ B+B+F+F+J+J=X \\ C+C+G+G+K+K=W \\ D+D+H+H+L+L=Y \end{matrix}\right\} Z+X+W+Y=60 \quad (1)$$

$$\left.\begin{matrix} a+a+e+e+i+i=z \\ b+b+f+f+j+j=x \\ c+c+g+g+k+k=w \\ d+d+h+h+l+l=y \end{matrix}\right\} z+x+w+y=60 \quad (2)$$

In these equations, the uppercase letters represent an individual's behavior under normal conditions, whereas the lowercase letters denote a person's actions in stressful situations. The resultant value for high-intensity situations provides an overall perspective of a person's behavior in tense situations, which may serve as a starting point for a subsequent analysis of an individual's conflict style.

The variables *Z*, *X*, *W*, and *Y* contain the total values obtained from both questionnaires in normal situations, while the variables *z*, *x*, *w*, and *y* indicate the same values in stressful situations. These variables are explained in further detail below:

> *Variable Z (and z)* contains the total value obtained from both questionnaires. This variable is related with the respondent's behavior in supporting other people.
>
> *Variable X (and x)* measures the degree to which a respondent is proactive.
>
> *Variable W (and w)* assesses the respondent's behavior in decision-making.
>
> *Variable Y (and y)* evaluates the degree to which a person is relaxed and agreeable.

Step 4: Define the person's activity level by following Rule 1 and using Equation 3.

Equation 3 and Rule 1 are used to determine the activity level of each person. Equation 3 is presented below:

$$M = Y + X; \qquad N = Z + W \qquad (3)$$

Where the variable M contains the values related with a person's positive and proactive attitudes and the variable N denotes the person's score as it relates to passive attitudes.

Accordingly, a person can be classified as an active or passive individual based on a comparison between these variables, which is known as Rule 1.

Rule 1: IF $M > N$ THEN a Person is Active
ELSE Person is Passive

Step 5: Define the person's orientation by using Rule 2 and Equation 4. Specifically, Equation 4 will determine the extent to which a person is people-orientated or task-orientated.

$$P = Z + Y; \qquad R = W + X \qquad (4)$$

Where P contains the results related with a person's tendency to support other people and R contains the score associated with the respondent's focus on task execution.

The resulting variables from Equation 4, P and R, can be compared to see which is greater. Accordingly, the respondent can be classified as "Oriented to persons" or "Oriented to tasks", as shown in Rule 2.

Rule 2: IF $P > R$ THEN a Person is Oriented to Persons
ELSE Person is Oriented to Tasks

Step 6: Use the following rules to determine the style for each person using the variables *X*, *Y*, *Z*, and *W* in a normal situation and *x*, *y*, *z*, and w in tense situation.

The variable $Diff_{i,j}$ identifies the difference between two variables, *i* and *j*, where $i, j \in [X,Y,Z,W, x, y, z, w]$. By analyzing the questionnaires' characteristics, it should be evident that the maximum difference between the two variables is 12. Thus Max $Diff_{i,j}=12$.

Furthermore,

- When the difference between i and j is equal to or greater than 80%, the difference is considered as a Remarkable Difference. For a Remarkable Difference to be evident, $Diff_{i,j} >= 10$.
- When the difference between i and j is between 50% and 80 %, it is considered as a Discrete Difference. For a Discrete Difference to exist, $Diff_{i,j} >= 6$ and $Diff_{i,j} <=9$.
- When the difference between i and j is less than or equal to 50% it is considered as a Short Difference. For a Short Difference to be evident, $Diff_{i,j} <= 5$.

Rule 1: IF $Diff_{i,j} >= 10$ THEN the person has a Dominant Style.

Rule 2: IF $Diff_{i,j} >= 6$ and $Diff_{i,j} <=9$, THEN the person has a Major/Minor Style.

Rule 3: IF $Diff_{i,j} <= 5$, THEN the person has a Mixed Style.

Rule 4.1: IF a person possesses the Mixed Style and the higher variables are X and Z, THEN the person has an "Administrative Mixed Style," which represents an organized and a responsible individual with the capabilities to resolve differences and overcome problems in difficult situations. Generally, team leaders and project managers possess such a personality style.

Rule 4.2: IF a person possesses the Mixed Style and the higher variables are *W* and *Z*, THEN the person has a "Technical Mixed Style," which represents a person that is calm, reasonable and honest. This type of individual usually does not take risks and follows secure and established traditions. People with technical mixed behavior should occupy roles such as architects, designers and analysts.

Rule 4.3: IF a person possesses the Mixed Style and the higher variables are *W* and *X*, THEN the person has an "Executive Mixed Style," which represents a person who promotes measures, results and metrics. This type of individual usually enjoys demonstrating results and progresses.

Rule 4.4: IF a person possesses the Mixed Style and the higher variables are *Y* and *X*, THEN the person has an “Energetic Mixed Style,” which represents a person who promotes activities, requires compensation, and recompenses other individuals. An individual with this style is generally optimistic and focused on satisfying the needs of other people. In particular, programmers and human resource managers possess this style.

Rule 4.5: IF a person possesses the Mixed Style and the higher variables are *Y* and *W*, THEN the person has a “Diplomatic Mixed Style.” This kind of person is usually friendly and humorous, and he/she performs the correct actions at appropriate times in an attempt to consistently please other people.

Rule 4.6: IF a person possesses the Mixed Style and the higher variables are *Y* and *W*, THEN the person has a “Developed Mixed Style,” which represents a responsible and an appreciative person. This individual is a good listener, promotes others and enjoys helping people attain their aspirations.

Step 7: After evaluating each individual in the preceding steps, describe the features of each worker based on the information contained in Table 3.

Table 3 Features to describe a person’s behavior

| **Collaborator (Z)** | **Promoter (Y)** | **Analyzer (W)** | **Controller (X)** |
|---|---|---|---|
| Idealistic,<br>Ambitious, and<br>Receptive<br>Loyal, Confident<br>Modest and<br>Attentive<br>Considered and<br>Collaborative<br>Courteous and<br>Responsive | Enthusiastic<br>and Energetic<br>Persuasive and<br>Motivational<br>Creative and<br>Positive<br>Optimistic and<br>Adaptable<br>Prudent and<br>Sensitive | Logical,<br>Practical,<br>Methodical<br>and Persistent<br>Efficient and<br>Careful<br>Judicious and<br>Reserved<br>Cautious and<br>Quiet | Strong and<br>Confident<br>Persistent,<br>Active and<br>Anxious<br>Quick to Act<br>Decisive and<br>Executive<br>Persuasive and<br>Imaginative<br>Entrepreneur |

Step 8: Complete Table 4 by inputting the descriptive information for all team members and use this information in the process of roles assignment.

Table 4 Resume table describes team members' features.

| Responsible | Normal situation | | | | Tense situation | | | |
|---|---|---|---|---|---|---|---|---|
| | Z | X | W | Y | z | x | w | y |
| Leader | | | | | | | | |
| Person 1 | | | | | | | | |
| Person 2 | | | | | | | | |
| Person … | | | | | | | | |
| Person N | | | | | | | | |
| Summary | | | | | | | | |

Once the descriptive information for all team members has been inputted into the table, the table can be analyzed to determine the degree of balance in the team. If there is a difference of more than 2 units among the columns, the team is not balanced and it requires some improvements.

An example of project member (X Person) results in this test is showing follows for a better understanding.

Results of Equations 1 and 2:

2+1+2+2+2+2=11-Z

3+4+4+4+4+4=23-X

4+2+1+1+3+3=14-W

1+3+3+3+1+1=12-Y

Results of Step 4 and Rule 1:

Y+X=M --- 12+23=35

Z+W=N --- 11+14=25 M>N

Results of Step 4 and Rule 2:

Z+Y=P --- 11+12=23

W+X=R --- 14+23=37 P<R

Through of responses analysis to the test was identified like an active person, its dominant feature is being controller. Others characteristics that can measure a person, according to the test, were a little farther from the main, which is closer is Analyzer. Its orientation is directed to tasks, such orientation is typical of directors, economic and mathematical.

Let it be a controller person reveals that at the time of confront a problem or a question whenever he believes have the solution and looking what is best. In stressful situations do not change their Controller characteristics, remain its key features.

Using the results of Equations 1 and 2 in Step 6, was obtained that person has Major/Minor style, this means that this person is Controller closely followed by a second feature, be Analyzer.

It can be concluded that this person is able to lead a team, take responsibility and challenges without fear because it has a strong self-confidence.

**Quality of life test**

Our quality of life test is based on the Chronic Heart Failure Questionnaire proposed by (Guyatt, et al., 1989). For our test, the questions have been divided into two categories: Fatigue (2, 4, 7 and 9) and Emotions (1, 3, 5, 6, 8, 10 and 11).

As demonstrated by the questionnaires in Table 5, each question has a rating of 1 to 7, where 1 indicates a lower quality of life and 7 denotes a higher quality of life. In each category, the scores for the questions are added together, as shown in Table 6. A low overall score indicates that a person's lifestyle causes unhappiness or frustration, whereas a higher score denotes that an individual's lifestyle does not have an adverse effect on that person. Quality of life questionnaires are often used to recommend that people experience more enjoyment in life (Rothstein & Goffin, 2006; ISQOLS, 1995).

Table 5 Quality of life questionnaire

| Question | Possible answer |
|---|---|
| Overall, during the last two weeks, how much of the time have you felt frustrated or impatient? | 1. All the time. 2. Most of the time.<br>3. A good amount of time. 4. Sometimes. 5. A little amount of time. 6. Hardly ever. 7. None at all. |
| How tired have you felt over the last two weeks? | 1. Extremely tired. 2. Very tired.<br>3. Quite tired.<br>4. Moderately tired. 5. Somewhat tired.<br>6. A little tired. 7. Not at all tired. |
| How often during the last two weeks have you felt inadequate, worthless or as | 1. All the time. 2. Most of the time.<br>3. A good amount of time.<br>4. Sometimes. |
| if you were a burden on others? | 5. A little amount of time.<br>6. Hardly ever. 7. None at all. |
| How much energetic have you felt in the last two weeks? | 1. Not at all. 2. A little bit.<br>3. Somewhat energetic.<br>4. Moderately energetic.<br>5. Quite energetic. 6. Very energetic.<br>7. Extremely energetic. |
| Overall, how much of the time did you feel upset, worried or depressed during the last two weeks? | 1. All the time. 2. Most of the time.<br>3. A good amount of time. 4. Sometimes. 5. A little amount of time.<br>6. Hardly ever. 7. None at all. |
| How much of the time during the last two weeks did you feel relaxed and free of tension? | 1. None at all. 2. Hardly ever.<br>3. A little amount of time.<br>4. Sometimes. 5. A good amount of the time. 6. Most of the times.<br>7. All the time. |
| How often during the last two weeks have you felt low in energy? | 1. All the time. 2. Most of the time.<br>3. A good amount of time. 4. Sometimes. 5. A little amount of time. 6. Hardly ever. 7. None at all. |
| In general, how often during the last two weeks have you felt discouraged or depressed? | 1. All the time. 2. Most of the time.<br>3. A good amount of time. 4. Sometimes. 5. A little amount of time. 6. Hardly ever. 7. None at all. |
| How often during the last two weeks have you felt worn out or sluggish? | 1. All the time. 2. Most of the time.<br>3. A good amount of time. 4. Sometimes. 5. A little amount of time. 6. Hardly ever. 7. None at all. |
| How happy, satisfied or pleased have you been with your personal life during the last two weeks? | 1. Not at all. 2. A little. 3. Somewhat. 4. Moderately happy. 5. Quite happy.<br>6. Very happy. 7. Extremely happy. |
| Overall, how often during the last two weeks have you | 1. All the time. 2. Most of the time.<br>3. A good amount of time. 4. Sometimes. |
| felt restless or tense? | 5. A little amount of time. 6. Hardly ever. 7. None at all. |

Table 6 Quality of life questionnaires - Range of values

| Category | Minimum score (worst function) | Maximum score (best function) |
|---|---|---|
| Fatigue | 4 | 28 |
| Emotional function | 7 | 49 |

**Roles Assignment Process**

There are two strategies for assigning specific roles to each person; first, roles could be assigned on the basis of an expert's judgment, or alternatively, automatic tools could recommend roles. Both strategies use information generated during the process of competence evaluation; specifically, the human resources organization chart and the results of the competence evaluation process are used as inputs in both of these strategies. However, regardless of which strategy is used, the system merely suggests roles for each person rather than assigning them to individuals on the basis of the applied algorithm. Using these two strategies, the system guarantees the use of the information obtained in the competence evaluation process, and it makes the following suggestions:

1. Each person should occupy a specific role in an appropriate workplace on the basis of his/her technical evaluation.
2. The teams should contain a balance in personalities. Specifically, a good team exists when the difference among variables X, Y, Z, and W is appropriate, indicating that the team members should have positive relations and work efficiently with one another.
3. In the first strategy of expert judgment, the team in charge of human resource acquisition should obtain the necessary information from each individual and assign the roles to each project member.
4. In the second strategy, the use of a semi-automatic tool or software helps to assign the roles by providing a preliminary structure of the human resources organization. However, this initial structure does not constitute the final human resources organization, which depends on the human resource acquisition team.
5. The semi-automatic strategy is not a substitute for human

experience. Accordingly, the results and suggestions generated by this strategy should leave room for modifications and adaptations by humans.

**Close Process**

Finally, the close process consists of two main activities:

1. The completion of the project staff
2. The communication of the acquisition results to the stakeholders.

Accordingly, we propose two reports to be generated in this process:

1. The report of human resource completion, which specifies the selected individuals and the position of each person in the project.
2. The report of acquisition processes, which includes all of the elements and aspects that were involved in the acquisition process.

**Experimental Results**

In order to verify the model's effectiveness, we selected six software development projects to which to apply the model and its tools. In April 2008 we applied the model and its tools to these projects. We proposed changes to these projects' organizational structure, which were applied in the following year. As explained in the following sections, our conclusions are based on three statistical tests.

**a) Projects characteristics**

All the projects analyzed were from a single center software product development. Project 1 was related to software quality management, and there were twelve people working on the project. In Project 2, twenty-two people were involved in a project addressing issues of business management. Project 3 concerned an e-commerce system, and the development team consisted of ten individuals. In Project 4, twenty-two people were working on various topics related to project management software systems. Project 5 involved the development of a statistical system with a twenty-three person team. Finally, Project 6 included twenty-two persons, and it focused on the development of a generic platform for conducting dynamic reports.

**Statistical tests to detect difficulties (April 2008)**

We applied the Wilcoxon Test, a non-parametric statistical method for the case of two related samples; to evaluate results of sociological test previously shown at point two referred to Competence Evaluation Process. In the test, we collected data for the following variables purposed in

sociological test:

- Activity level (active or passive),
- Expected Role in Normal Situations
- Current Role in Normal Situations
- Expected Role in Tense Situations
- Current Role in Tense Situations
- State (normal or tense),
- Project balance (yes or no)
- Individual orientation (task-oriented or people-oriented)

The purpose for including these variables was to measure their status in teams that had been created without taking into account the elements suggested in the research. Assess the level of each of them, propose changes as suggested by the proposal and then assess whether there had been improvements in productivity and personal relationships of these teams.

A *simple random sample* method was used to select two of the six projects, representing 33% of the original project sample. These two projects, Projects 5 and 6, were evaluated in October 2009, and the results demonstrated that the application of our model significantly improved the projects' performance. Specifically, we detected considerable differences in these projects on the basis of the following variables:

- Orientation of Normal Situations and Tense Situations: Human beings change their orientation significantly depending on whether the situation is considered normal or tense. In normal situations, the orientation of team members is focused on individuals, whereas in tense situations, team members are focused more strongly on the task.
- Expected Role vs. Current Role (Normal Situations): There is a remarkable difference between the expected role and the current role in normal situations.
- Expected Role vs. Current Role (Tense Situations): There is a significant difference between the expected role and the current role in high-pressure situations.

**Recommended changes for improving projects' performance**

Our observations demonstrated that most individuals were performing tasks that were different than those recommended by the model. Accordingly, their roles within the team required modification, and, in order to improve the project performance, we recommended the reallocation of human resources within the project, the most convenient rearrangement of roles, and a reorganization of the projects to achieve more balanced teams.

#### Statistical test, checking the improvements

In order to evaluate the projects, we compared the results of Projects 5 and 6 with their previous results. Specifically, two pairs of samples (Project-5 2008, Project-5 2009) and (Project-6 2008, Project-6 2009), were compared, focusing on the teams' balance and performance. After the application of corrective measures, Projects 5 and 6 were observed to have balanced teams. We applied the Wilcoxon Test to compare the results, which are displayed in Table 8. This table reflects the increased productivity of Projects 5 and 6 after the application of the corrective actions to the projects' teams.

Table 8 Software requirements, time and productivity of projects

| Projects | Requirements (R) | Time (Month) | Productivity (R/M) |
|---|---|---|---|
| Project 5 April 2008 | 49 | 16 | 3.06 |
| Project 5 October 2009 | 94 | 13 | 7.23 |
| Project 6 April 2008 | 55 | 7 | 7.86 |
| Project 6 October 2009 | 89 | 11 | 8.09 |

To estimate productivity was taken into account the efficiency (E) of the teams taking as indicator the number of requirements (R) between the time, in months needed to developing them (Oficina Nacional de Normalización, 2007). Variables such as the number of people in the teams, the characteristics of the requirements specification and the daily time utilized, remained stable throughout the experiment so were not taken into account for calculating efficiency.

$$E = R \div months \quad (5)$$

#### Threats to validity

There are two major threats to the model´s validity:

1. The quality of collected data depends on the tests application. The organization must assure the quality of the questionnaires application required for attitudes and for life assessment.
2. The results obtained in this study could be influenced by other factors

like the improvement of the individual competence during the project development.

## Conclusions

The acquisition process of human resource management consists of four main activities: initialization, competence evaluation, roles assignment, and communications to the stakeholders.

Within the stage of competence evaluation, we have proposed three types of tests: the technical test, the quality of life test, and the psychological-sociological test. These tests form the basis for our proposed model for evaluating the quality and balance of software teams, which has been applied to real software projects. The experimental results demonstrate that the application of the model improved the productivity of project teams.

We have also proposed two alternatives to the role assignment of individuals: manual techniques and automatic techniques; however, algorithms do not substitute for human experience, as they need to be revised by humans.